\documentclass[aps,pre,reprint, amsmath, amssymb,twocolumn]{revtex4}

\usepackage[amssymb]{SIunits}
\usepackage{bm}
\usepackage[retainorgcmds]{IEEEtrantools}
\usepackage{graphicx}
\usepackage{mathrsfs}
\usepackage{amsmath}
\usepackage{amssymb}

\newcommand{\ud}{\,\mathrm{d}}

\DeclareMathOperator{\tr}{tr}

\newcommand{\trans}{^{\text{T}}}
\newcommand{\inv}{^{-1}}
\newcommand{\PP}{\mathcal{P}}
\newcommand{\entropy}{\mathcal{S}}
\newcommand{\ir}{^\text{irr}}
\newcommand{\rev}{^\text{rev}}
\newcommand{\diag}{\text{diag}}

\begin{document}

\title{Entropy production in linear Langevin systems}
\date{\today}
\author{Gabriel. T. Landi, T\^ania Tom\'e and M\'ario J. de Oliveira}
\affiliation{Instituto de F\'isica da Universidade de S\~ao Paulo,  05314-970 S\~ao Paulo, Brazil}

\begin{abstract}

We study the entropy production rate in systems described by linear Langevin equations, containing  mixed even and odd variables under time reversal. Exact formulas are derived for several important quantities in terms only of the means and covariances of the random variables in question. These include the total rate of change of the entropy, the entropy production rate, the entropy flux rate and the three components of the entropy production. All equations are cast in a way suitable for large-scale analysis of linear Langevin systems. Our results are also applied to different types of electrical circuits, which suitably illustrate the most relevant aspects of the problem. 

\end{abstract}
\maketitle{}

%
%
\section{\label{sec:int}Introduction}
%
%

Non-equilibrium systems have been the subject of intensive research for several decades. This is partly motivated  by  its broad range of applications in, e.g.,  physics, chemistry and biology. However, and most importantly, further progress in these areas is still hampered by more fundamental questions. Unlike equilibrium statistical mechanics, which is by now a well established theoretical framework, in non-equilibrium statistical mechanics several questions remain unanswered. Particularly challenging is the microscopic definition of \emph{entropy production}. For, in the context of non-equilibrium thermodynamics \cite{DeGroot1961}, it provides the pathway through which  irreversibility  is described. 

Suppose a certain system undergoes a change of state, from state $A$ to state $B$. If done reversibly, the total change in the entropy $\mathcal{S}$ of the system is $\Delta \mathcal{S} = \int \ud Q/T$, where $T$ is the temperature and $Q$ is the heat poured into the system. Hence $\int \ud Q/T$ is defined as minus the entropy flux from the system to the environment. If the process is irreversible we have instead $\Delta \mathcal{S} \geq \int \ud Q/T$. The difference, being a positive quantity, is called the entropy production $\mathfrak{P}$
\[ 
\Delta \mathcal{S} - \int  \frac{\ud Q}{T} = \mathfrak{P} \geq0
\]
 It is customary to divide by $\Delta t$ and write, instead
\begin{equation}\label{eq:dsdt_def}
\frac{\ud \mathcal{S}}{\ud t} = \Pi(t) - \Phi(t)
\end{equation}
In this equation $\Pi(t)$ is the entropy production rate of the system, which is always nonnegative,  and $\Phi(t)$ is the entropy flux rate \emph{from} the system \emph{to} the environment. For systems in a non-equilibrium steady-state (NESS) we have $\ud \entropy/\ud t = 0$, which implies 
\[
\Pi_0 = \Phi_0 \geq 0
\]
It is only in  thermodynamic equilibrium that the equality in this equation is expected to hold.

Traditionally, non-equilibrium thermodynamics was founded on the basis of conservation equations. Nowadays, however, it has been realised that there are several advantages in using stochastic processes instead. For instance, by comparing forward and backward experiments, it enables one to the relate the entropy  production directly to the stochastic trajectories of the system   \cite{Esposito2010,Verley2012,Spinney2012,Crooks2000}.  When describing a non-equilibrium system in terms of stochastic processes, much of the  focus has naturally been on Markovian dynamics, in particular using the master equation \cite{Ford2012,Tome2012,Germany1976,Verley2012,Esposito2010} or the Fokker-Planck  approach \cite{Tome2010,VandenBroeck2010,Spinney2012}, which will be the choice for the present paper. We also note that non-Markovian dynamics have also been  recently investigated \cite{Garcia-Garcia2012}.  
 
  
Several formulas for the entropy production rate have been derived for both representations. In all cases, however,  these are  written in terms of integrals involving probability currents (cf. Sec.~\ref{sec:ent_gen}). Thus, they are not easily computed in most situations. In this paper we will focus on linear Langevin systems; i.e., where all terms appearing in the stochastic differential equations are linear in the independent variables. First of all, one must always emphasise the importance of linear systems, in view of the many circumstances  in which they appear in nature. Moreover, we will show  that, for such systems, it is possible to obtain exact formulas for the entropy production rate in terms of  the means and variances of the independent variables. This enables one to study more complex situations, which are prohibitive for non-linear systems. In fact, with the scheme to be derived below, it is possible to implement extremely efficient numerical procedures to study the entropy production even in large-scale systems. 

The entropy production gives insight into the properties of systems out of equilibrium. And with the formulas developed in this paper, it becomes simple to compute the entropy production even for the most complex linear systems. Moreover, these results are not restricted to the steady-state as in most recent papers, but also naturally include the time-dependence. We thus hope that these results are of value to deepen our understanding of non-equilibrium physics. 

When discussing entropy production, it is  paramount to distinguish between variables that are even and odd under time reversal. With this in mind, we shall illustrate our results by applying them  to electrical circuits. These are excellent platforms for this type of problem,  since they contain mixed  even (charge and voltage) and odd (current) variables. These studies trace  back to the works of Landauer \cite{Landauer1975}, and the discussion about the differences between the minimum entropy principle and the maximum entropy principle. They were also  recently discussed in  Ref.~\cite{Bruers2007} using the  Onsager-Machlup Lagrangian  \cite{Onsager1953}. 

The problem and the basic underlying equations will be stated in Sec.~\ref{sec:prob}. General remarks about the entropy production rate will be given in Sec.~\ref{sec:ent_gen} and the derivation of the formulas for linear systems will be carried out in Sec.~\ref{sec:ent_lin}. Some of the lengthier calculations were postponed to the appendix. The applications in electrical circuits are contained in Sec.~\ref{sec:app} and the conclusions in Sec.~\ref{sec:conc}.

With the exception of Sec.~\ref{sec:app}, we shall try to maintain the following notation. Vectors are denoted by lower-case letters, such as $x = (x_1,\ldots,x_n)$ and matrices by upper-case letters such as $A$ and $B$. The exception will be random variables, such as $X$, which will also be in upper-case letters. 
All vectors  are treated as column vectors, with $x\trans$ representing the corresponding row vector.
Scalars and scalar functions are denoted by calligraphic letters such as $\mathcal{S}$ or $\mathcal{P}$ (the only exception being $\Pi$ and $\Phi$ in Eq.~(\ref{eq:dsdt_def})). The gradient of $\mathcal{P}$ with respect to $x$ is abbreviated as $\partial \mathcal{P}/\partial x$.

%
%
\section{\label{sec:prob}Systems described by Langevin equations}
%
%

\subsection{\label{ssec:prob_lang}The Langevin equation}

Let $X = (X_1, \ldots,X_n)$ be a vector of random variables satisfying 
\begin{equation}\label{eq:lin_1}
\dot{X} = f(X,t) + B \dot{\xi}(t)
\end{equation}
In this equation $f(x,t)$ is an arbitrary $n$-dimensional function of $x$ and the time, t. The $\xi(t)$ are $m$ independent \emph{standard} Wiener processes and, therefore,   $B$ is an $n\times m$ matrix \footnote{Choosing $B$ to be $n\times m$ is convenient because it includes the possibilities that a variable contain more than one source of noise, or that the same noise is shared with more than one variable.}. In this paper we shall  focus mainly on linear systems, for which we write
\begin{equation}\label{eq:func_f}
f(x,t) = - A x + b(t),
\end{equation}
where $A$ is $n\times n$ and $b(t)$ is an arbitrary n-dimensional function of time. We shall also make the reasonable assumption that all eigenvalues of $A$ are in the open right-plane, thence guaranteeing the stability of the solutions.

The expectation of $x$ is denoted by  $\bar{x} = \langle X \rangle$, both  notations being used interchangeably.  The equation for the time evolution of $\bar{x}$ is obtained directly by taking the expectation of  Eq.~(\ref{eq:lin_1}):
\begin{equation}\label{eq:lin_ave}
\frac{\ud \bar{x}}{\ud t} = f(\bar{x}, t) = - A \bar{x} + b(t)
\end{equation} 

Next we obtain the equation describing the time evolution of the second moments. The latter can be constructed from the outer product $\langle X X\trans\rangle$, which gives a matrix whose $(i,j)$-th entry is $\langle X_i X_j \rangle$. The result -- obtained for instance by  discretising time, taking the outer product and then averaging  -- is 
\begin{equation}\label{eq:second_moms}
\frac{\ud \langle X X\trans\rangle}{\ud t} = X f\trans + f X\trans + B B\trans
\end{equation}
For linear systems it is more convenient to work with the covariance matrix:
\begin{equation}\label{eq:cov_mat}
\Theta = \langle X X\trans \rangle - \langle X \rangle \langle X \rangle \trans
\end{equation}
Note that $\Theta$ is symmetric by construction. Moreover, being a covariance matrix, it is also positive definite. 
Using Eqs.~(\ref{eq:lin_ave}) and (\ref{eq:second_moms}) and assuming linearity, as in Eq.~(\ref{eq:func_f}), we find that 
\begin{equation}\label{eq:lyap_diff}
\frac{\ud \Theta }{\ud t} = - (A\Theta  + \Theta A\trans) + 2 D
\end{equation}
which is a matrix differential equation giving the time evolution of the covariance matrix. Here $D$ is the $n\times n$ diffusion tensor defined as 
\begin{equation}\label{eq:diffusion_tensor}
D = \frac{1}{2} B B\trans
\end{equation}
If $B$ has full row rank, then $D$ is positive-definite; otherwise it is positive semi-definite. 


Eq.~(\ref{eq:lyap_diff}) shows an important property of linear systems of Langevin equations, which is seldom discussed in the literature: all terms involving the external forcing term, $b(t)$, dropped out. This means that the variability of the random variables in question is not influenced by external forces. In other words, we could say linear systems are not amenable to \emph{synchronisation}.

If $b(t) = b$, a constant, then the system will eventually reach equilibrium (we are making the explicit assumption that $A$ is stable). The equilibrium value of  $\bar{x}$ is read immediately from Eq.~(\ref{eq:lin_ave}): $\bar{x}_0 = A\inv b$. Similarly, setting $\dot{\Theta}_0 = 0$ in Eq.~(\ref{eq:lyap_diff}) we obtain the matrix equation
\begin{equation}\label{eq:lyap}
A\Theta_0  + \Theta_0 A\trans = 2D
\end{equation}
This is a very important equation. It appears frequently in the literature of electrical engineering and control systems where it goes by the name of \emph{continuous time Lyapunov equation}. It is also a particular case of the broader class of  \emph{Sylvester equations}.

It seems appropriate to stop now to briefly discuss the solution methods of Eq.~(\ref{eq:lyap}). For algebraic solutions we can transform it into a linear system of equations as follows. Let $A\otimes B$ denote the Kronecker product of two matrices $A$ and $B$ and define the operation $\text{vec}(A)$ as that of creating a vector by stacking the columns of $A$. Then Eq.~(\ref{eq:lyap}) can be written as 
\begin{equation}\label{eq:lyap_kron}
\Big[ (I \otimes A) + (A\otimes I)\Big] \text{vec}(\Theta_0) = 2\text{vec}(D)
\end{equation}
Given that $\Theta$ is symmetric, several equations will be repeated and, in the event that $A$ is sparse, several equations will be of the form $0=0$, thence simplifying somewhat the computations. On the other hand, this approach should \emph{never} be used for numerical calculations. The computational complexity of Eq.~(\ref{eq:lyap_kron}) is $O(n)^6$. However, specialised algorithms have been developed which reduce this to $O(n)^3$, a substantial improvement \cite{Bartels1972}.

\subsection{The Fokker-Planck equation}

Let  $\mathcal{P} (x,t)$ denote the probability density function (PDF) corresponding to the vector of random variables $X$. The Fokker-Planck equation for $\mathcal{P}(x,t)$ associated with the general Langevin equation~(\ref{eq:lin_1}) reads
\begin{equation}\label{eq:FP1}
\frac{\partial \mathcal{P}}{\partial t} = - \sum\limits_j \frac{\partial }{\partial x_j} \Big[ f_j(x,t) \mathcal{P}\Big] + \sum\limits_{j,k} \frac{\partial^2}{\partial x_j \partial x_k} \Big[ D_{jk} \mathcal{P} \Big]
\end{equation}
It is also convenient to write this equation in the form of a continuity equation. For this let us define the probability current 
\begin{equation}\label{eq:g}
g(x,t) = f(x,t) \PP(x,t) - D \frac{\partial \PP(x,t)}{\partial x}   
\end{equation}
Eq.~(\ref{eq:FP1}) can then be written as 
\begin{equation}\label{eq:FP2}
\frac{\partial \mathcal{P}}{\partial t}  = - \frac{\partial }{\partial x} \cdot g 
\end{equation}
where the operation in the right-hand side denote the divergence of $g$.

The Fokker-Planck equation for linear systems satisfies the very important property that it's solution must be of the form of a multivariate normal distribution \footnote{This can be understood intuitively by noting that, upon discretising time in Eq.~(\ref{eq:lin_1}), $x_{t+\Delta t}$ is formed by summing normally distributed random variables ($\Delta \xi$ and $x_{t}$). But sums of normally distributed random variables must also be normally distributed and so will  $x_{t+\Delta t}$.}:
\begin{equation}\label{eq:FP_sol}
\mathcal{P} (x,t) = \frac{1}{\sqrt{(2\pi)^n | \Theta|}} \exp\Big\{ -\frac{1}{2} (x-\bar{x})\trans \Theta\inv (x-\bar{x})\Big\}
\end{equation}
where $\bar{x}$ and $\Theta$ are given by Eqs.~(\ref{eq:lin_ave}) and (\ref{eq:lyap_diff}) (both being, in general, functions of time, which has been omitted for clarity). Here $|\Theta|$ denotes the determinant. 

In the event that $b(t)$ is a constant, the system will reach a steady-state whose distribution is
\begin{equation}\label{eq:FP_sol_ss}
\mathcal{P}_0 (x) = \frac{1}{\sqrt{(2\pi)^n | \Theta_0|}} \exp\Big\{ -\frac{1}{2} (x-\bar{x}_0)\trans \Theta_0\inv (x-\bar{x}_0)\Big\}
\end{equation}
where $\bar{x}_0 = A\inv b$ and $\Theta_0$ is given by Eq.~(\ref{eq:lyap}).

Throughout this paper we will make frequent use of the fact that both $\PP(x,t)$ and $\partial \PP/\partial x$ vanish at the boundaries of the probability space, so that cross terms appearing when integrating by parts can always be neglected. While this assumption is reasonable in most systems, it is in fact entirely justified for the linear problem, given that the PDF is described by a multivariate normal distribution, as in Eq.~(\ref{eq:FP_sol}).

\subsection{Distinction between variables that are even and odd  under time reversal}

In studying the entropy production, it is important to distinguish between odd variables, which reverse sign under time reversal, and even variables, which do not. Examples of even variables include the position in mechanical systems and charges or voltages in circuits, their odd counterparts being velocities and currents. 
Following \cite{Spinney2012} and \cite{Risken1989} let us define, for each variable $x_i$, a quantity $\epsilon_i$ such that $\epsilon_i = \pm 1$ if $x_i$ is even or odd respectively. Moreover, let us define a diagonal matrix $E = \diag(\epsilon_1,\ldots, \epsilon_n)$. Then, time reversal is achieved by the operation $x\to Ex$ \footnote{The matrix $E$ satisfies $E\inv = E$. Moreover, while this will not be necessary in the present paper, it is interesting to note that the operators $\frac{1}{2}(I\pm E)$ may be used to select the even and odd variables respectively}. 

Let us also divide $f(x,t)$ in Eq.~(\ref{eq:lin_1}) into irreversible and reversible parts:
\begin{equation}\label{eq:f_div}
f(x,t) = f\ir(x,t) + f\rev(x,t)
\end{equation}
where 
\begin{IEEEeqnarray}{rCl}\label{eq:f_ir_rev}
f\ir(x,t) & = & \frac{1}{2} \left[ f(x,t) + E f (E x,t)\right] = E f\ir (E x,t)  \nonumber \\
&& \\
f\rev(x,t) & = & \frac{1}{2} \left[ f(x,t) - E f (E x,t)\right] = -  E f\rev(E x,t) \nonumber 
\end{IEEEeqnarray}

It is convenient to separate even and odd variables by writing $x = (x_1,x_2)$, where it is agreed that $x_1$ contains all even variables and  $x_2$ all odd ones (the dimensions of $x_1$ and $x_2$ depending on the problem in question). We then have that
\begin{equation}\label{eq:timerev}
E x = (x_1, -x_2)
\end{equation}
Let us also define 
\begin{equation}\label{eq:A_blocks}
A = \begin{bmatrix}
	A_{11}  &  A_{12}  \\[0.2cm]
	A_{21}  &  A_{22} 
\end{bmatrix}
\end{equation}
where the block matrices $A_{ij}$ have dimensions compatible with those of $x_1$ and $x_2$. 
Then, according to Eq.~(\ref{eq:f_ir_rev}), we may write
\begin{equation}\label{eq:A_div}
A = A\ir + A\rev
\end{equation}
where
\begin{equation}\label{eq:Air}
A\ir = \begin{bmatrix}
	A_{11}  & 0  \\[0.2cm]
	0  &  A_{22} 
\end{bmatrix},
\qquad
A\rev = \begin{bmatrix}
	0           &  A_{12}  \\[0.2cm]
	A_{21} & 0 
\end{bmatrix}
\end{equation}
Finally, we divide the external forcing term, $b(t)$ as $b(t) = (b_1 (t), b_2(t))$ with $b_1$ and $b_2$ having the same dimensions as $x_1$ and $x_2$. It then follows again from Eq.~(\ref{eq:f_ir_rev}) that
\begin{equation}\label{eq:b_ir_rev} 
b\ir = (b_1,0),\quad \quad b\rev = (0,b_2)
\end{equation}
Explicit examples of this separation are given in Sec.~\ref{sec:app} for several types of electric circuits. 

It will also be important to distinguish the irreversible and reversible parts of the probability current:
\begin{equation}\label{eq:g_div}
g = g\ir + g\rev
\end{equation}
where
\begin{IEEEeqnarray}{rCl}
g\ir(x,t) &=&  f\ir(x,t) \PP(x,t) - D \frac{\partial \PP(x,t)}{\partial x}   \label{eq:gir}   \\[0.2cm]
g\rev(x,t) &=&  f\rev(x,t) \PP(x,t)   \label{eq:grev}   
\end{IEEEeqnarray}

\subsection{Conditions for equilibrium}

In the steady-state $\dot{\mathcal{P}}_0 = 0$ and thence $\frac{\partial}{\partial x} \cdot g_0 =0$. Moreover, the probability currents $g_0(x)$ should be such that even components change sign under time reversal whereas odd components do not; i.e., 
\begin{equation}
g_0(x) = - E g_0(Ex)
\end{equation}
Using the definitions in Eq.~(\ref{eq:g_div})-(\ref{eq:grev}) we find
\[
g_0(Ex) = \Big[f\rev(Ex)  + f\ir(Ex) \Big]\mathcal{P}_0(Ex) - D \frac{\partial \mathcal{P}_0(Ex)}{\partial Ex}
\]
With the aid of Eq.~(\ref{eq:f_ir_rev}) this may be written as 
\[
g_0(Ex) = E \Big[-f\rev(x)+f\ir(x)\Big]\mathcal{P}_0(Ex) - DE \frac{\partial \mathcal{P}_0(Ex)}{\partial x}
\]
Whence, 
\begin{IEEEeqnarray}{rCl}\label{eq:g_condition}
g_0(x)+Eg_0(Ex) &=& f\rev(x)\Big[\mathcal{P}_0(x) -\mathcal{P}_0(Ex) \Big] \nonumber \\[0.2cm] &+& f\ir (x) \Big[\mathcal{P}_0(x) + \mathcal{P}_0(Ex) \Big]   \\[0.2cm]
&-&  D \frac{\partial \mathcal{P}_0(x)}{\partial x} - EDE \frac{\partial \mathcal{P}_0(Ex)}{\partial x} \nonumber \\[0.2cm]
&=& 0 \nonumber
\end{IEEEeqnarray}
If $\mathcal{P}_0(Ex) = \mathcal{P}_0(x)$ (the distribution is an even function of the odd variables), the first term in Eq.~(\ref{eq:g_condition}) vanishes. Moreover, if $EDE = D$ (which is generally true if $D$ is diagonal or block-diagonal), then by comparing Eq.~(\ref{eq:g_condition}) with the definition (\ref{eq:gir}) we conclude that
\[
 g_0\ir\equiv f\ir (x) \mathcal{P}_0(x) - D \frac{\partial \mathcal{P}_0(x)}{\partial x}  = 0 
\]
i.e., the irreversible portion of the current is zero in the steady state. This condition has already been discussed in the context where $D$ is diagonal \cite{Risken1989}. For the more general case  presented here we see that the two conditions for this to happen are that $\mathcal{P}_0(Ex) = \mathcal{P}_0(x)$ and $EDE = D$. 

Being this the case, we find from $\frac{\partial}{\partial x} \cdot g_0 =0$ that 
\[
\frac{\partial}{\partial x} \cdot g_0\rev  = \left[\frac{\partial }{\partial x} \cdot f\rev(x) \right] \mathcal{P}_0 + [f\rev(x)]\trans \frac{\partial \mathcal{P}_0}{\partial x} = 0 
\]
In most physical systems the reversible part of the force, $f\rev(x)$, is divergence-less:
\begin{equation}\label{eq:rev_div}
\frac{\partial}{\partial x} \cdot f\rev(x) = 0
\end{equation}
 In this case, we may use Eq.~(\ref{eq:gir}) to write $\frac{\partial \mathcal{P}_0}{\partial x} = D\inv f\ir (x) \mathcal{P}_0(x)$ so as to we finally arrive at 
\begin{equation}\label{eq:equilibrium}
[f\rev(x)]\trans D\inv [f\ir(x)] = 0
\end{equation}
This is the required condition for equilibrium. It says that the vectors $f\rev$ and $f\ir$ should be orthogonal in a space whose metric is $D\inv$. In the event that $D$ is semi-definite, then by $D\inv$ we mean the pseudo-inverse of $D$.

%
%
\section{\label{sec:ent_gen}Entropy production: general statement}
%
%

The following calculations follow closely those of Ref.~\cite{Tome2010}, but generalise it in two ways. First, it includes systems with arbitrary combinations of even and odd variables and second, it includes non-diagonal $D$'s.
 
The entropy is defined as 
\begin{equation}\label{eq:entropy_def}
\mathcal{S} = - \int \mathcal{P} \log \mathcal{P} \ud x
\end{equation}
Differentiating with respect to time, using  Eq.~(\ref{eq:FP2}) and integrating  by parts we find 
\begin{equation}\label{eq:dsdt_general}
\frac{\ud \entropy}{\ud t} = - \int \frac{\partial \PP}{\partial t} \log \PP \ud x = -\int g\trans \frac{\partial \PP}{\partial x}    \frac{\ud x}{\PP}
\end{equation}
Separating $g$ as in Eq.~(\ref{eq:g_div}) we find for the term involving $g\rev$, after integrating by parts:
\[
 -\int (g\rev)\trans \frac{\partial \PP}{\partial x}    \frac{\ud x}{\PP}  =  \int \left[\frac{\partial }{\partial x} \cdot f\rev\right]  \PP\ud x = 0
\]
since we are assuming $f\rev$ is divergence-less [cf. Eq.~(\ref{eq:rev_div})]. Hence, Eq.~(\ref{eq:dsdt_general}) becomes 
\begin{equation}\label{eq:dsdt_withgir}
\frac{\ud \entropy}{\ud t} = -\int {g\ir}\trans \frac{\partial \PP}{\partial x}    \frac{\ud x}{\PP}
\end{equation}
Next we use Eq.~(\ref{eq:gir}) to write 
\[
\frac{\partial \PP}{\partial x}    =  D\inv \big[f\ir(x,t) \PP - g\ir(x,t)\big]
\]
and obtain 
\begin{equation}\label{eq:dsdt_2parts}
\frac{\ud \entropy}{\ud t} = \int {g\ir}\trans D\inv g\ir \frac{\ud x}{\PP} - \int {g\ir}\trans D\inv f\ir \ud x
\end{equation}
The first term is a quadratic form. Since $D$ is positive definite, it is always non-negative and can thence be identified with the entropy production rate:
\begin{equation}\label{eq:PI}
\Pi(t) = \int {g\ir}\trans D\inv g\ir \frac{\ud x}{\PP} 
\end{equation}
Consequently, the entropy flux rate is found to be 
\begin{equation}\label{eq:PHI}
\Phi(t) = - \int {g\ir}\trans D\inv f\ir \ud x
\end{equation}

Using a different approach, Eq.~(\ref{eq:PI}) was further separated in  Ref.~\cite{Spinney2012} into three terms. These, unlike Eq.~(\ref{eq:PI}), hold \emph{only} for  diagonal $D$. They also  assume that a steady-state, time-independent, configuration exists. Usually, we will simply take this to mean that $b(t) = b$, a constant.
In our present notation, the formulas are:
\begin{IEEEeqnarray}{rCl}
\Pi_1(t) &=& - \int \frac{\partial \PP}{\partial t} \log\frac{\PP(x,t)}{\PP_0(x)}\ud x 
\label{eq:PI1} \\[0.2cm]
\Pi_2(t) &=& \int [g_0\ir(Ex)]\trans D\inv  [g_0\ir(Ex)]\trans  \frac{\PP(x,t)}{[\PP_0(Ex)]^2} \ud x 
\IEEEeqnarraynumspace
\label{eq:PI2} \\[0.2cm]
\Pi_3(t) &=& - \int \frac{\partial \PP}{\partial t} \log\frac{\PP_0(x)}{\PP_0(Ex)}\ud x 
\label{eq:PI3}
\end{IEEEeqnarray}
where $\PP_0$ is given by Eq.~(\ref{eq:FP_sol_ss}) and $g_0\ir$ is obtained from $g\ir$ in Eq.~(\ref{eq:gir}) by replacing $\PP$ with $\PP_0$. 
The first contribution,  termed nonadiabatic, is related to the relaxation processes of the system and should be zero at a NESS (non-equilibrium steady state). On the other hand, the second contribution, termed adiabatic, is directly related to the absence of detailed balance in the system and will be the sole contribution to $\Pi$ when a NESS is reached. Both $\Pi_1$ and $\Pi_2$ have been shown \cite{Spinney2012} to obey integral fluctuation theorems. However, the same is not true of the third term, $\Pi_3$.

%
%
\section{\label{sec:ent_lin}Entropy production for linear systems}
%
%

Last section summarised a series of results, valid for general Langevin systems. In this section we shall specialise them to linear systems, as in Eq.~(\ref{eq:func_f}), and obtain formulas for all relevant quantities. The major advantage of considering linear systems is that all results can be expressed in terms of $\bar{x}$ and $\Theta$, thus avoiding the necessity of performing the several integrals involved in those expressions. 

\subsection{Rate of the change of the total entropy}

We begin with the total rate of change of the entropy. Substituting Eq.~(\ref{eq:FP_sol}) for $\log \mathcal{P}$ in Eq.~(\ref{eq:entropy_def}) we find 
\[
\mathcal{S} =\frac{1}{2} \int \Big[  (x-\bar{x})\trans \Theta\inv (x- \bar{x}) + \log (2\pi)^n |\Theta|\Big] \mathcal{P} \ud x
\]
The last term is simply $\log (2\pi)^n |\Theta|$. To compute the first term we proceed as follows. Since $\Theta$ is symmetric positive definite, we can construct it's square, triangular Cholesky factorisation $\Theta = Q Q\trans$. Now consider the transformation $x = Q z + \bar{x}$. It transforms
$(x-\bar{x})\trans \Theta\inv (x- \bar{x}) = z\trans z$. Whence, we see that this transformation produces a standard multivariate normal, where all variables are statistically independent. Hence, the first term acquires the form $\langle z\trans z \rangle = n$. We thus obtain 
\begin{equation}\label{eq:ent_lin}
\mathcal{S}(t) = \frac{1}{2} \log |\Theta(t)| + \frac{n}{2} \log (2\pi e)
\end{equation}

Differentiating Eq.~(\ref{eq:ent_lin}) with respect to time and using a known formula of matrix calculus yields
\begin{equation}\label{eq:dsdt}
\frac{\ud \mathcal{S}}{\ud t} = \frac{1}{2} \tr\left(\Theta\inv \frac{\ud \Theta}{\ud t}\right)
\end{equation}
From Eq.~(\ref{eq:lyap}) we have 
\[
\Theta\inv \frac{\ud \Theta}{\ud t} = -\Big[ \Theta\inv A \Theta + A\trans \Big] +2 \Theta\inv D
\]
The matrices $A\trans$ and $\Theta\inv A \Theta $ are similar, and thence share the same trace. Thus, Eq.~(\ref{eq:dsdt}) becomes
\begin{equation}\label{eq:dsdt_lin}
\frac{\ud \mathcal{S}}{\ud t} = \tr(\Theta\inv D - A)
\end{equation}
where $\Theta\inv$ is generally a function of time. This is the required formula. It gives the rate of change of the total entropy of the system, which is seen to be a function of only the inverse of the covariance matrix, $\Theta\inv$. Note, as before, that $\dot{S}$ is entirely independent of  $b(t)$ (the same will not be true for $\Pi$ and $\Phi$). 

In the steady-state we set $\dot{\mathcal{S}} = 0$ in Eq.~(\ref{eq:dsdt_lin}) to find
\begin{equation}\label{eq:ent_equi}
\tr(\Theta_0\inv D - A) = 0
\end{equation}
This relation can also be derived directly from the steady-state solution, Eq.~(\ref{eq:lyap}), by multiplying both sides by $\Theta_0\inv$ and then taking the trace.

%
\subsection{Entropy production rate and entropy flux rate}
%

We now derive formulas for $\Pi$ and $\Phi$. The formulas for $\Pi_1$, $\Pi_2$ and $\Pi_3$ are derived by similar methods in the Appendix.
Before we proceed, let us establish a convenient mathematical operation which will be used extensively, specially in the appendix. We will frequently encounter quantities of the following form:
\[
\int (F x + u)\trans (G x + v) \PP(x,t) \ud x
\]
which correspond to an average over $\PP(x,t)$ of a quadratic form; here $F$, $G$, $u$ and $v$ are arbitrary matrices and vectors. When we expand the product, there will be only one term that is of second order in $x$, which will have the form $x\trans F\trans G x$. Taking the expectation of this term yields
\[
\langle x\trans F\trans G  x \rangle = \sum\limits_{ij} (F\trans G)_{ij} \langle x_i x_j \rangle
\]
Next we use Eq.~(\ref{eq:cov_mat}) to write $\langle x_i x_j \rangle = \Theta_{ij} + \bar{x}_i \bar{x}_j$, which results in
\[
\langle x\trans F\trans G x \rangle = \tr (F\trans G \Theta) + \bar{x}\trans F\trans G \bar{x}
\]
The last term is again a quadratic form and may thus be reincorporated into the original product. We thus arrive at the following relation:
\begin{IEEEeqnarray}{rCl}\label{eq:ave_method}
\int (F x + u)\trans (G x + v) \PP(x,t) \ud x &= & \tr (F\trans G \Theta) \\[0.2cm]
&+& (F \bar{x} + u)\trans (G \bar{x} + v)\nonumber
\end{IEEEeqnarray}
That is, whenever we average a quadratic form, there will be a trace term containing $\Theta$ and the original term inside the integral, with $x$ replaced by $\bar{x}$.

With this result, it is now straightforward to compute the entropy production rate in Eq.~(\ref{eq:PI}).
From Eq.~(\ref{eq:FP_sol}) we have 
\begin{equation}\label{eq:dpdx}
\frac{\partial \PP}{\partial x} = - \Theta\inv (x - \bar{x}) \PP(x,t)
\end{equation}
Using this in Eq.~(\ref{eq:gir}) we find
\begin{IEEEeqnarray}{rCl}\label{eq:glin}
g\ir &=& (-A\ir x + b\ir) \PP - D \frac{\partial \PP}{\partial x}\nonumber\\[0.2cm]
   &=&  \Big[(D\Theta\inv - A\ir) x + b\ir - D\Theta\inv \bar{x}\Big] \PP \\[0.2cm]
   &=& (F x + u) \PP \nonumber
\end{IEEEeqnarray}
where we have defined $F = D\Theta\inv -A\ir$ and $u = b\ir - D\Theta\inv \bar{x}$. 
By applying the method described in Eq.~(\ref{eq:ave_method}) we obtain
\begin{IEEEeqnarray*}{rCl}
\Pi(t) &=& \int (Fx+u)\trans D\inv (Fx+u) \PP \ud x \\[0.2cm]
         &=& \tr(F\trans D\inv F \Theta) + (F\bar{x} + u)\trans D\inv (F\bar{x}+u)
\end{IEEEeqnarray*}
Using that  $F\bar{x} + u$ = $b\ir - A\ir \bar{x}$ and expanding the matrix product inside the trace we finally find that 
\begin{IEEEeqnarray}{rCl}\label{eq:PI_lin}
\Pi(t) &=& \tr(D\Theta\inv - A\ir) + \tr({A\ir}\trans D\inv A\ir \Theta - A\ir) \nonumber \\[0.2cm]
 &+& (A\ir \bar{x} - b\ir)\trans D\inv (A\ir \bar{x} - b\ir)
\end{IEEEeqnarray}
which is the required result. 

We can identify the first term in Eq.~(\ref{eq:PI_lin}) as being simply $\ud \entropy/\ud t$ in Eq.~(\ref{eq:dsdt_lin}). Hence, from Eq.~(\ref{eq:dsdt_def}) we immediately find 
\begin{IEEEeqnarray}{rCl}\label{eq:PHI_lin}
\Phi(t) &=&  \tr({A\ir}\trans D\inv A\ir \Theta - A\ir) \nonumber \\[0.2cm]
 &+& (A\ir \bar{x} - b\ir)\trans D\inv (A\ir \bar{x} - b\ir)
\end{IEEEeqnarray}
This result can also be easily derived from Eq.~(\ref{eq:PHI}) using the same method. 

In the steady-state, making use of Eq.~(\ref{eq:ent_equi}) we find
\begin{IEEEeqnarray}{rCl}\label{eq:PI_NESS}
\Pi_0 = \Phi_0 &=&  \tr({A\ir}\trans D\inv A\ir \Theta_0 - A\ir)  \nonumber \\[0.2cm]
&+& (A\ir \bar{x}_0 - b\ir)\trans D\inv (A\ir \bar{x}_0 - b\ir)
\end{IEEEeqnarray}

We can also use the same approach to derive formulas for the three contributions to the entropy production rate, defined in Eqs.~(\ref{eq:PI1})-(\ref{eq:PI3}). These calculations, however, are somewhat lengthier. Thus, we will simply state the results here  and postpone the details to the Appendix. We emphasise, once again, that these results are valid only for diagonal $D$ and constant $b$. 

First, the result for $\Pi_1(t)$ is:
\begin{IEEEeqnarray}{rCl}\label{eq:PI1_lin}
\Pi_1(t) &=& \tr( D \Theta\inv - A) +  \tr(A\trans \Theta_0\inv \Theta - A) \nonumber \\[0.2cm]
             &+& (\bar{x} - \bar{x}_0)\trans \Theta_0\inv D \Theta_0\inv (\bar{x}-\bar{x}_0)
\end{IEEEeqnarray}
The nature of $\Pi_1$, as being related to the relaxation of the system, becomes quite visible from the structure of this formula. 
Next, we obtain for $\Pi_2$:
\begin{IEEEeqnarray}{rCl}\label{eq:PI2_lin}
\Pi_2(t) &=& \tr({A\ir}\trans D\inv A\ir \Theta) - \tr(A\trans E \Theta_0\inv E \Theta)  \nonumber \\[0.2cm]
	    &+& (A\ir \bar{x} - b\ir)\trans D\inv (A\ir \bar{x} - b\ir) \\[0.2cm]
	    &-& (A\bar{x}-b)\trans E \Theta_0\inv (E\bar{x} - \bar{x}_0)\nonumber
\end{IEEEeqnarray}
Finally, the result for $\Pi_3$ is:
\begin{IEEEeqnarray}{rCl}\label{eq:PI3_lin}
\Pi_3(t) &=& \tr(A\trans E \Theta_0\inv E \Theta - A) - \tr(A\trans \Theta_0\inv \Theta - A) \nonumber \\[0.2cm]
	    &+& (A\bar{x}-b)\trans E\Theta_0\inv (E\bar{x}-\bar{x}_0) \\[0.2cm]
	     &-& (\bar{x} - \bar{x}_0)\trans \Theta_0\inv D \Theta_0\inv (\bar{x}-\bar{x}_0)
\end{IEEEeqnarray}

%
%
\section{\label{sec:app}Application to electrical circuits}
%
%

Henceforth we will soften our previous notation of vectors and matrices, and use the  usual nomenclature of electrical circuits. We shall consider electrical circuits connected to different heat baths. The coupling constant is, in this case, the resistance $R$. The fluctuations will be assumed to correspond to white noise with spectral density given by the Jonhson-Nyquist formula $\sqrt{2 R T}$, where $T$ is the temperature of the bath. 

\subsection{\label{ssec:app_rl}RL circuit in series}

We begin by considering a simple, yet very instructive, example of a resistor $R$ and an inductor $L$ in series with a battery providing a constant emf $\mathcal{E}$. The independent variable is the current $I$ through the resistor, which is odd under time reversal. The equation for the time evolution of $I$ is obtained from Kirchhoff's voltage law:
\begin{equation}\label{eq:rl_basic}
\frac{\ud I}{\ud t} = - \frac{R}{L} I + \frac{\mathcal{E}}{L} + \sqrt{\frac{2RT}{L^2}} \dot{\xi}
\end{equation}
Making reference to the notation of Sec.~\ref{ssec:prob_lang} (all matrices are now $1\times1$), we have $A = R/L$, $b = \mathcal{E}/L$,  $B = \sqrt{2RT/L^2}$ and $D = (1/2) BB\trans = RT/L^2$. For simplicity, we will assume $I(t=0) = 0$. 
The expectation of $I$ as a function of time is obtained by solving Eq.~(\ref{eq:lin_ave}):
\begin{equation}\label{eq:rl_mean}
\bar{I}(t) = \frac{\mathcal{E}}{R} (1 - e^{-R t/L})
\end{equation}
The steady-state current is  clearly seen to be $\bar{I}_0 = \mathcal{E}{R}$. 
The covariance ``matrix'' is $\Theta = \langle I^2 \rangle - \bar{I}^2$ and  satisfies the differential equation  [cf. Eq~(\ref{eq:lyap_diff})]
\begin{equation}\label{eq:rl_covdif}
\frac{\ud \Theta}{\ud t} = - \frac{2R}{L} \Theta + \frac{2RT}{L^2}
\end{equation}
whose solution is 
\begin{equation}\label{eq:rl_cov}
\Theta(t) = \frac{T}{L} (1 - e^{-2 Rt/L})
\end{equation}

The current is an odd variable under time reversal. Thence, we have from Eq.~(\ref{eq:Air}) that $A\ir = A = R/L$ and, from Eq.~(\ref{eq:b_ir_rev}),  $b\ir = 0$. With this in mind, the entropy production rate is readily computed from Eq.~(\ref{eq:PI_lin}):
\begin{equation}\label{eq:rl_pi}
\Pi(t) = \frac{\mathcal{E}^2}{RT} (1-e^{-Rt/L})^2 + \frac{R}{L} \frac{e^{-2 Rt/L}}{e^{2Rt/L} - 1}
\end{equation}
If we take the limit $t\to \infty$ we obtain for the NESS entropy production rate:
\begin{equation}\label{eq:rl_pi0}
\Pi_0 = \Phi_0 = \frac{\mathcal{E}^2}{RT} = \frac{R \bar{I}_0^2}{T}
\end{equation}
This formula shows that, in the NESS, there is a constant production of entropy due to the dissipation in the resistor.  

\begin{figure}
\centering
\includegraphics[width=0.45\textwidth]{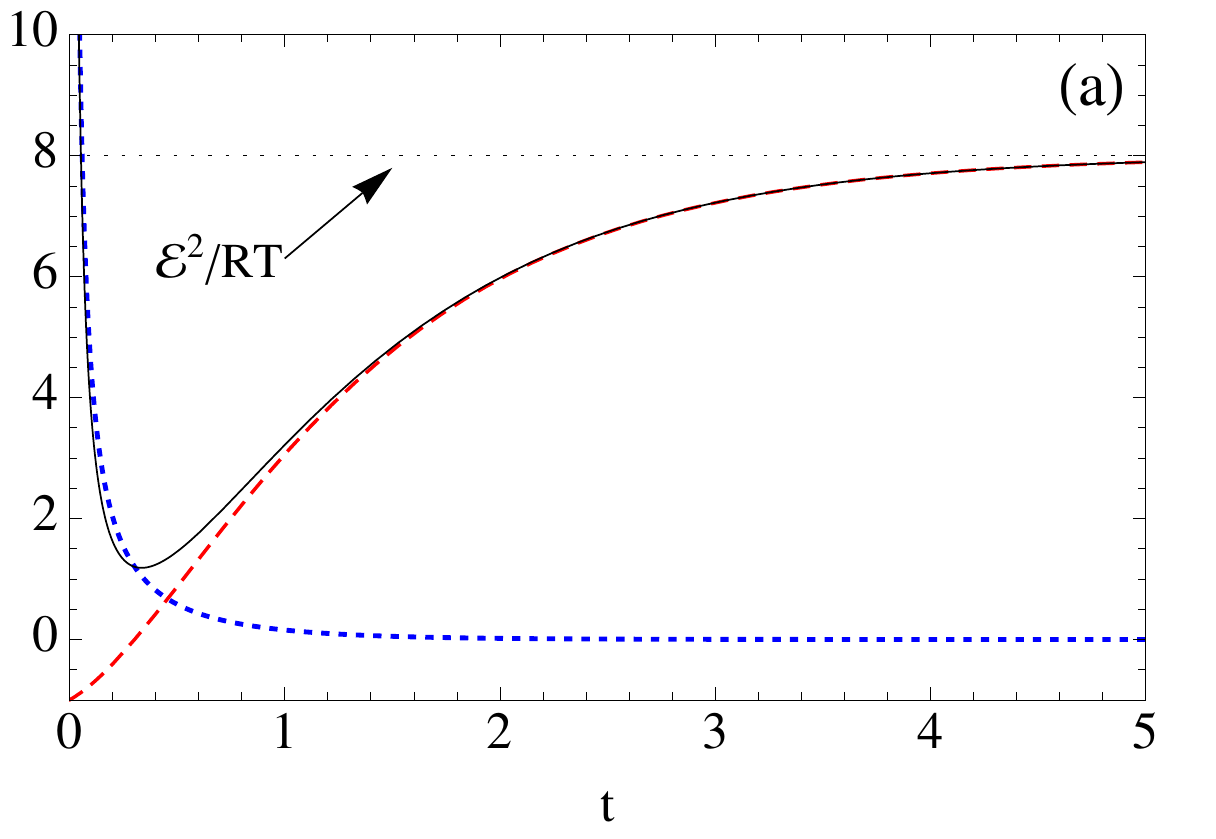}
\includegraphics[width=0.45\textwidth]{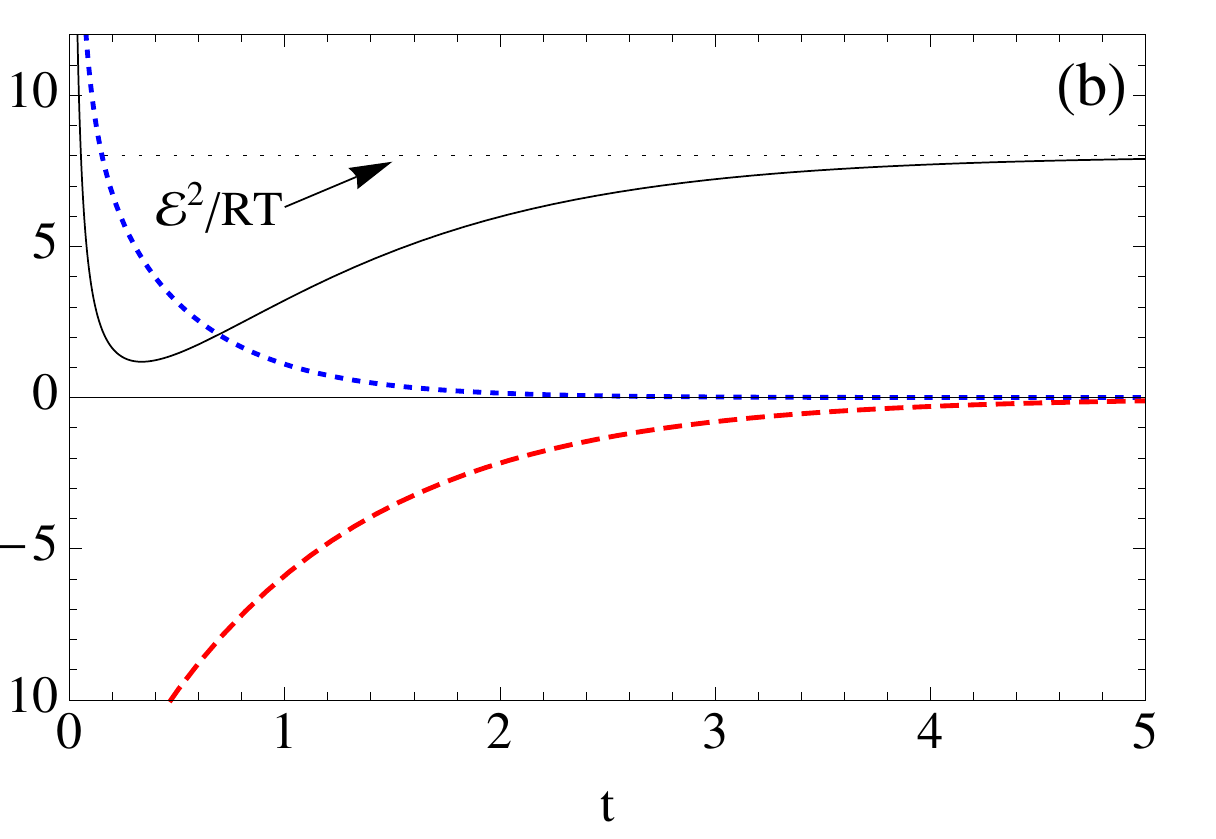}
\caption{\label{fig:rl_ent}Entropy production as a function of time for an RL circuit with $\mathcal{E} = 2~\volt$, $R = 1~\ohm$, $L = 1~\henry$ and $k_B T = 1/2~\joule$. (a) $\ud \entropy/\ud t$ (blue, dotted), $\Phi(t)$ (red, dashed) and $\Pi(t)$ (black); the dotted line indicate the NESS value $\Pi_0 = \Pi_2 = \mathcal{E}^2/RT$. (b) The different contributions to the entropy production rate;  $\Pi_1$ (blue, dotted), $\Pi_3$ (red, dashed) and $\Pi$ (black). Again, $\Pi_2$ is given by the dotted horizontal line.}
\end{figure}

The formula for the total rate of change of the entropy is
\[
\frac{\ud \entropy}{\ud t} = \frac{R}{L} \frac{1}{e^{2Rt/L} - 1}
\]
The entropy flux rate is then simply $\Phi = \Pi - \ud\entropy/\ud t$. These quantities are illustrated in  Fig.~\ref{fig:rl_ent}(a) for arbitrary values of the parameters. 

If we had, instead, studied an RC circuit then the independent variable would be the voltage through the resistor, which is even under time reversal. In this case, the formulas would be quite similar, except that now, the emf would be a part of $b\ir$. The resulting expression for the entropy production would now go to zero as $t\to\infty$ since, in an RC circuit, the current (which is responsible for the dissipation) goes to zero as $t\to\infty$. 

Continuing further, we may now compute formulas for the three parts of the entropy production rate, Eqs.~(\ref{eq:PI1_lin})-(\ref{eq:PI3_lin}):
\begin{IEEEeqnarray*}{rCl}
\Pi_1(t) &=& \frac{\mathcal{E}^2}{RT} e^{-2Rt/L} + \frac{R}{L} \frac{e^{-2 Rt/L}}{e^{2Rt/L} - 1} \\[0.2cm]
\Pi_2(t) &=& \frac{\mathcal{E}^2}{RT} \\[0.2cm]
\Pi_3(t) &=& -2 \frac{\mathcal{E}^2}{RT} e^{-Rt/L}
\end{IEEEeqnarray*}
These formulas are physically very rich. Note that $\Pi_2$, which is exactly the contribution related to the absence of detailed balance, is simply the steady-state entropy production rate $\Pi_0$ in Eq.~(\ref{eq:rl_pi0}).   Graphs for these quantities are illustrated in Fig.~\ref{fig:rl_ent}(b).

\subsection{\label{ssec:app_RR}Mixed RC and RL circuits}

\begin{figure}
\centering
\includegraphics[width=0.3\textwidth]{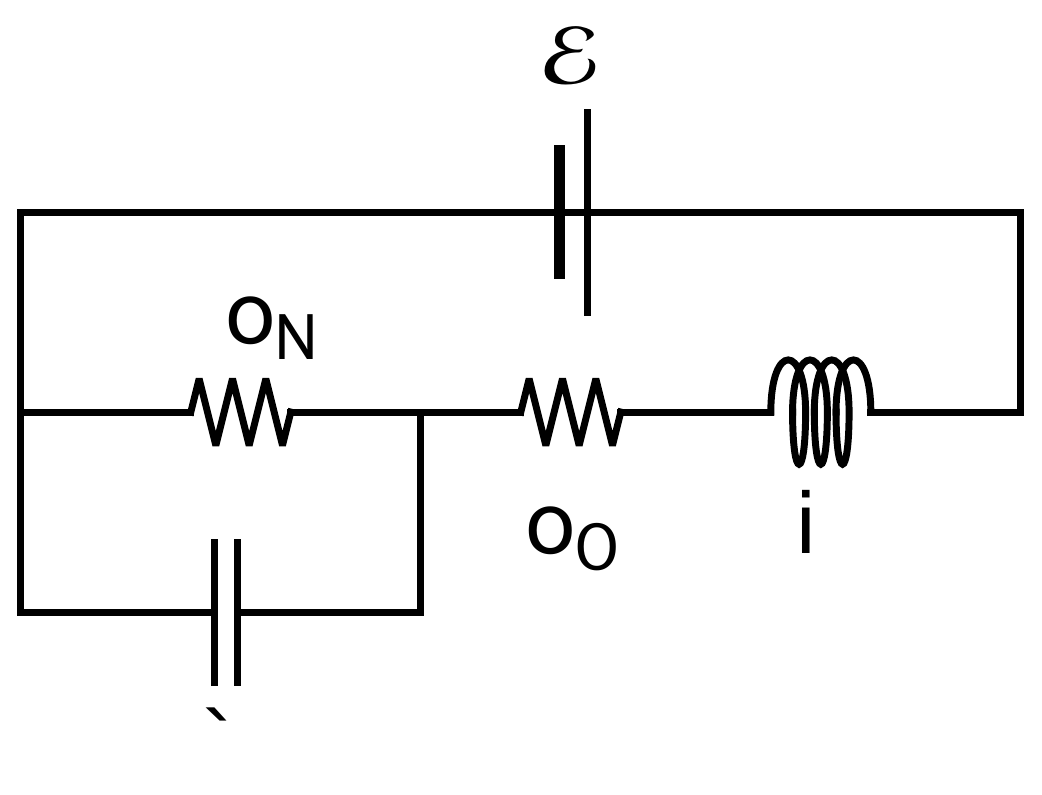}
\caption{\label{fig:circuit}Scheme of the circuit studied in Sec.~\ref{ssec:app_RR}. Each resistor is connected to a heat bath at a different temperature.}
\end{figure}

Let us now turn to the circuit denoted in Fig.~\ref{fig:circuit}, which was also studied in Ref.~\cite{Bruers2007}. We now have, as independent variables, the voltage $U$ through resistor $R_1$ and the current $I$ through the resistor $R_2$; i.e., we have  mixed even and odd variables. Using Kirchhoff's voltage law we find the equations
\begin{IEEEeqnarray}{rCl}\label{eq:rr_basic}
\dot{U} &=& \frac{I}{C} - \frac{U}{ R_1 C} + \sqrt{\frac{2 T_1}{R_1 C^2}} \dot{\xi}_1 \nonumber \\
& & \\
\dot{I} &=& \frac{\mathcal{E} - U - R_2 I}{L} + \sqrt{\frac{2 R_2 T_2}{L^2}}\dot{\xi_2} 
\end{IEEEeqnarray}

In our usual matrix notation we have from Eq.~(\ref{eq:Air}): 
\[
A\ir = \begin{bmatrix}
 1/CR_1 & 0 \\
 0 & R_2/L 
 \end{bmatrix},\qquad 
A\rev = \begin{bmatrix}
 0 & -1/C \\
 1/L & 0
 \end{bmatrix}
 \]
 and from Eq.~(\ref{eq:b_ir_rev})
 \[
 b\ir = 0, \qquad b\rev = b = \begin{bmatrix} 0 \\ \mathcal{E}/L \end{bmatrix}
 \]
The diffusion matrix, Eq.~(\ref{eq:diffusion_tensor}), is 
\[
D = \begin{bmatrix}
T_1 /R_1 C^2 & 0 \\
0 & R_2 T_2/L^2
\end{bmatrix}
\]

The formulas for the time-dependence of the entropy production rates are now somewhat cumbersome. Thus, we will illustrate their time dependence numerically and provide  formulas only for the steady-state quantities. The equilibrium values of $U$ and $I$ are 
\[
\bar{U}_0 = \frac{R_1}{R_1+R_2} \mathcal{E},\qquad
\bar{I}_0 = \frac{1}{R_1+R_2} \mathcal{E}
\]
The steady-state values of the covariance matrix are obtained by solving Eq.~(\ref{eq:lyap}) (the notation Var and Cov meaning variance and covariance respectively):
\begin{IEEEeqnarray*}{rCl}
\text{Var}(U) &=& \frac{\alpha}{C} \Big[L(R_1 + R_2) T_1 + C R_1 R_2 (R_2 T_1 + R_1 T_2)\Big] \\[0.2cm]
\text{Var}(I) &=& \frac{\alpha}{L} \Big[L(R_1T_1 + R_2 T_2) + C R_1 R_2 (R_1 + R_2) T_2\Big] \\[0.2cm]
\text{Cov}(U,I) &=& \alpha \Big[R_1 R_2 (T_2 - T_1)\Big] \\[0.2cm]
\alpha &:=& [(R_1+R_2)(L+C R_1 R_2)]^{-1}
\end{IEEEeqnarray*}

From these results we may now compute the steady-state entropy production rate from Eq.~(\ref{eq:PI_NESS}):
\begin{equation}\label{eq:mixed_Pi0}
\Pi_0 = \Phi_0 = \frac{\bar{U}_0}{R_1 T_1} + \frac{R_2 \bar{I}_0}{T_2} +\frac{R_1 R_2 (T_2-T_1)^2}{(R_1+R_2) (L + C R_1 R_2) T_1 T_2}
\end{equation}
The structure of this result is worth noticing. The first two terms are similar to those appearing in Eq.~(\ref{eq:rl_pi0}), each one corresponding to a resistor. They stem from the absence of detailed-balance, related to the presence of the batteries. 
In addition to them, however, there is now a second term related to the lack of detailed balance due to the presence of two heat baths. We see that it depends only on $(T_1-T_2)^2$. It is thus zero when both temperatures are equal, and shows that it is irrelevant which temperature is the largest.

It is worth comparing Eq.~(\ref{eq:mixed_Pi0}) with a similar result obtained for the same circuit in Ref.~\cite{Bruers2007}. There, the third term was not present. This follows because, in the formalism of the Onsager-Machlup Lagrangian \cite{Onsager1953}, the entropy production rate is related to the extremum of the Lagrangian with respect to the independent variables. Since the last term in Eq.~(\ref{eq:mixed_Pi0}) is independent of $U$ and $I$, it does not appear in that formalism. 

\begin{figure}
\centering
\includegraphics[width=0.45\textwidth]{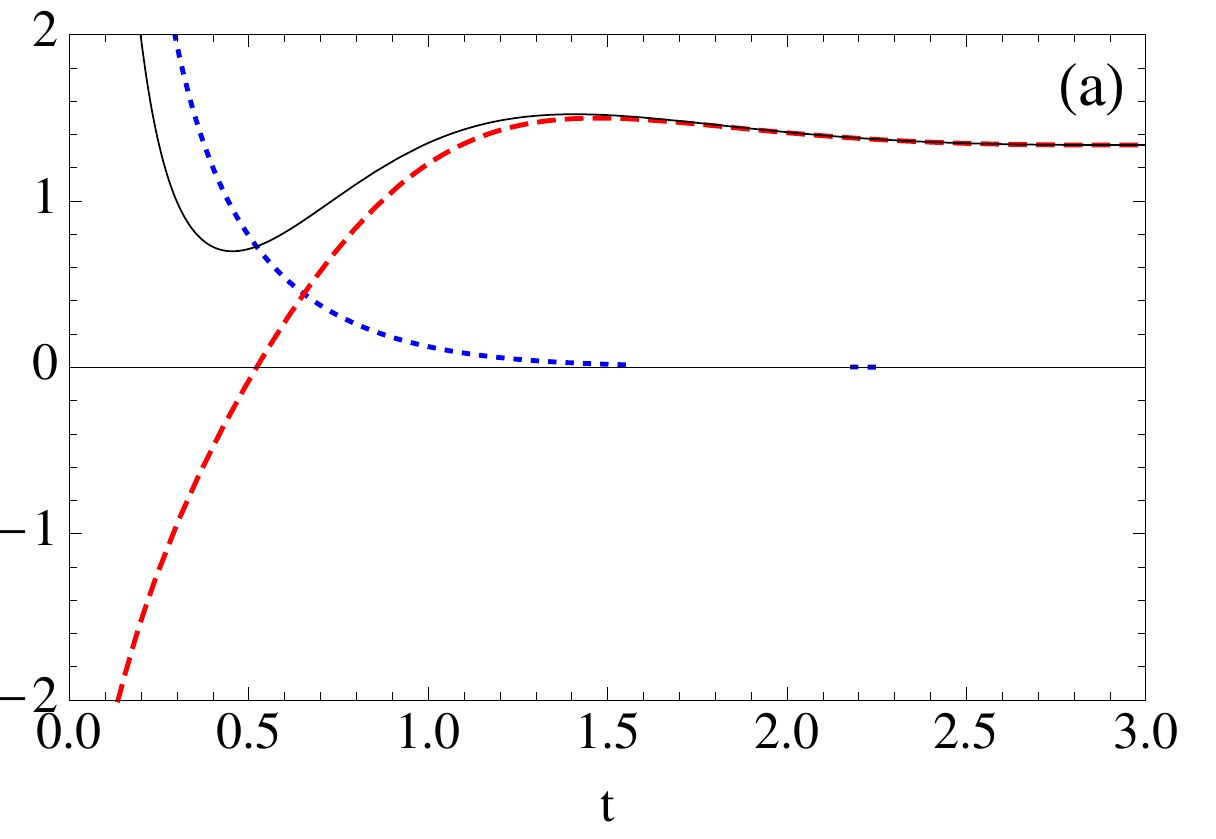}
\includegraphics[width=0.45\textwidth]{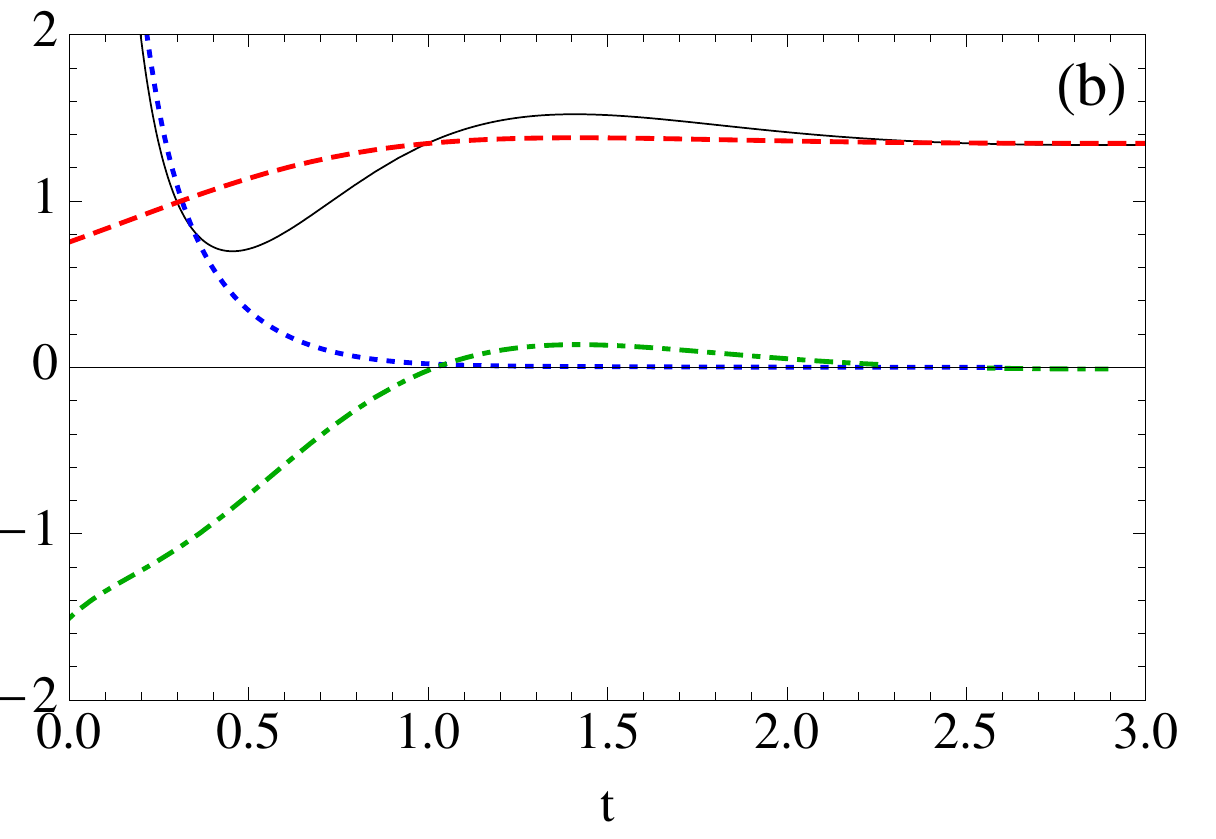}
\caption{\label{fig:mixed_ent}Entropy production as a function of time for the circuit in Fig.~\ref{fig:circuit}, with $\mathcal{E} = 2~\volt$, $R = 2~\ohm$, $R_2 = 1~\ohm$, $L = 1~\henry$, $C = 0.2~\farad$, $k_B T_1 = 1~\joule$ and $k_B T_2 = 2~\joule$. (a) $\ud \entropy/\ud t$ (blue, dotted), $\Phi(t)$ (red, dashed) and $\Pi(t)$ (black) (b) The different contributions to the entropy production rate;  $\Pi_1$ (blue, dotted), $\Pi_2$ (red, dashed) and $\Pi_3$ (green, dash-dotted); $\Pi(t)$ is again shown in black.}
\end{figure}

In Fig.~\ref{fig:mixed_ent}(a) we show the total rate of change of the entropy, the entropy production rate and the entropy flux rate as a function of time, for arbitrary choices of the parameters. Fig.~\ref{fig:mixed_ent}(b) then shows the 3 contributions to the entropy production rate.

\subsection{\label{ssec:app_2rl}Inductively coupled RL circuits}

Next, we turn to the interesting example of two inductively coupled RL circuits, each connected to a different heat bath. Let the suffixes $1$ and $2$ denote the quantities pertaining to the two circuits and let $m$ denote the mutual inductance. Then, the corresponding dynamical equations are 
\begin{IEEEeqnarray*}{rCl}
L_1 \dot{I}_1 + m     \dot{I}_2 &=& - R_1 I_1 + \mathcal{E}_1 + \sqrt{2R_1 T_1} \dot{\xi}_1 \\[0.2cm]
m    \dot{I}_1 + L_2  \dot{I}_2 &=& - R_2 I_2 + \mathcal{E}_2 + \sqrt{2R_2 T_2} \dot{\xi}_1 \\[0.2cm]
\end{IEEEeqnarray*}
Let us defined the matrix
\[
M = \begin{bmatrix}
	L_1 & m \\
	m   & L_2
\end{bmatrix}
\]
Then, in the notation of Eq.~(\ref{eq:lin_1}) we  have 
\[
A = M\inv \begin{bmatrix}
	 R_1  &  0 \\
	0       &   R_2
\end{bmatrix}
\]	
and
\[
D = M\inv 
\begin{bmatrix}
	 R_1 T_1  &  0 \\
	0       &   R_2 T_2
\end{bmatrix}
M\inv
\]
In this case, we see that the diffusion matrix $D$ is not diagonal, so formulas (\ref{eq:PI1_lin})-(\ref{eq:PI3_lin}) are note applicable. We will  thus restrict our discussion to the steady-state expression for the entropy production rate. The steady-state currents are simply 
\[
\bar{I}_{10} = \frac{\mathcal{E}_1}{R_1},\qquad \bar{I}_{20} = \frac{\mathcal{E}_2}{R_2}
\]
The steady-state covariance matrix is computed from Eq.~(\ref{eq:lyap}). It reads
\[
\Theta_0 = \begin{bmatrix}
	T_1 		&		\frac{m R_2(T_1-T_2)}{L_2 R_1 + L_1 R_2} \\[0.2cm]
	\frac{m R_1(T_2-T_1)}{L_2 R_1 + L_1 R_2}	& T_2 
\end{bmatrix} M\inv
\]	

On noting that both variables are odd under time reversal, we obtain from Eq.~(\ref{eq:PI_NESS}):
\begin{IEEEeqnarray}{rCl}\label{eq:2rl_pi0}
\Pi_0 &=& \frac{\mathcal{E}_1^2}{R_1 T_1} + \frac{\mathcal{E}_2^2}{R_2 T_2}\\[0.3cm]
          &+& \frac{m^2 R_1 R_2}{(L_1L_2 - m^2)(L_2R_1+L_1R_2)} \frac{(T_1-T_2)^2}{T_1 T_2} 
\end{IEEEeqnarray}
Again, we arrive at a structure compatible with the two sources of disorder: the first two terms come from the batteries and the last from the presence of two heat baths. As expected, the last term goes to zero when $m\to0$ since the mutual inductance is the sole mechanism of coupling in the system.

Finally, it is also illustrative to consider three inductively coupled RL circuits. For simplicity, we will make the assumption that all mutual inductances and self-inductances are  the same, being $m$ and $L$ respectively. The result is then 
\begin{IEEEeqnarray*}{rCl}
\Pi_0 &=& \frac{\mathcal{E}_1^2}{R_1 T_1} + \frac{\mathcal{E}_2^2}{R_2 T_2} + \frac{\mathcal{E}_3^2}{R_3 T_3} + \frac{m^2}{W } \Big[\alpha_1 T_1 (T_2-T_3)^2 
\\[0.3cm]
&+&  \alpha_2 T_2 (T_1-T_3)^2 + \alpha_3 T_3 (T_1-T_2)^2 \Big] \frac{1}{T_1 T_2 T_3}
\end{IEEEeqnarray*}
where
\begin{IEEEeqnarray*}{rCl}
W &=& (L-m)(L+2m) \Big[ 2 m^2 R_1 R_2 R_3 \\[0.2cm]
&+&  L (L+m)(R_1 + R_2) (R_2+R_3)(R_1+R_3) \Big]
\end{IEEEeqnarray*}
and 
\[
\alpha_1 = R_2 R_3 \Big[2 m R_1^2 + L (R_1+R_2)(R_2+R_3)\Big]
\]
with similar expressions for $\alpha_2$ and $\alpha_3$. Again, the results depend only on the temperature differences, as expected.

%
%
\section{\label{sec:conc}Conclusions}
%
%

In conclusion, we have studied the entropy production rate in systems of linear Langevin equations. Linearity and it's consequences enabled us to compute formulas for (i) the total rate of change of the entropy, (ii) the entropy production rate \emph{per se}, (iii) the entropy flux rate and (iv) the three contributions to the entropy production stemming from different physical sources. All these formulas were expressed in terms of the mean and covariance matrix of the random variables in questions. This makes their implementation in large-scale systems quite easy. Our results were applied to electrical circuits of various types. For instance, we have shown that two circuits which are  coupled only via their mutual inductance have a steady-state entropy production rate related to the difference in temperature between the two circuits. 

\acknowledgements 

The authors acknowledge the funding agency FAPESP for the financial support. 

%
%
\appendix
\section{The three parts of the entropy production rate}
%
%

In this section we describe the steps to obtain formulas~(\ref{eq:PI1_lin})-(\ref{eq:PI3_lin}) from Eqs.~(\ref{eq:PI1})-(\ref{eq:PI3}). 

\subsection{Equation for $\Pi_1(t)$}

We begin with $\Pi_1$ in Eq.~(\ref{eq:PI1}), which can be written as 
\begin{equation}\label{eq:PI1_tmp}
\Pi_1(t) = \frac{\ud \entropy}{\ud t} + \Gamma_1
\end{equation}
where $\ud \entropy/\ud t$ is given by Eq.~(\ref{eq:dsdt_lin}) and
\begin{equation}\label{eq:gamma1}
\Gamma_1(t) = \int \frac{\partial \PP}{\partial t} \log \PP_0(x) \ud x
\end{equation}
Using the FP Eq.~(\ref{eq:FP2}) and integrating by parts we find
\[
\Gamma_1 = \int g\trans \frac{\partial \PP_0(x)}{\partial x} \frac{\ud x}{\PP_0(x)}
\]
Similarly to Eq.~(\ref{eq:dpdx}) we have $\partial \PP_0(x) /\partial x = - \Theta_0\inv (x-\bar{x}_0) \PP_0(x)$. Using also Eq.~(\ref{eq:glin}), but with $g$ instead of $g\ir$; i.e.  with $F = D\Theta\inv - A$ and $u = b - D\Theta\inv \bar{x}$ we find 
\[
\Gamma_1 =- \int (Fx+u)\trans \Theta_0\inv (x-\bar{x}_0) P(x,t) \ud x
\]
Using the method of Eq.~(\ref{eq:ave_method}) we obtain
\[
\Gamma_1 = - \tr(F\trans \Theta_0 \Theta) - (F\bar{x} + u)\trans \Theta_0 (\bar{x}-\bar{x}_0)
\]
This can be simplified to read
\begin{equation}\label{eq:gamma1_lin}
\Gamma_1(t) = \tr(A\trans \Theta_0\inv \Theta - A) + (A\bar{x} - b)\trans \Theta_0\inv (\bar{x}-\bar{x}_0)
\end{equation}

The last term in Eq.~(\ref{eq:gamma1_lin}) can also be modified as follows. Multiplying both sides of Eq.~(\ref{eq:lyap}) by $\Theta_0\inv$ we find 
\begin{equation}\label{eq:lyap_inv}
\Theta_0\inv A + A\trans \Theta_0\inv = 2 \Theta_0\inv D \Theta_0 \inv
\end{equation}
Now let $y$ be an arbitrary vector. Noting that $y\trans M y = y\trans M\trans y$ we find 
\begin{IEEEeqnarray*}{rCl}
y\trans \Theta_0\inv A y + y\trans A\trans \Theta_0\inv y &=& 2 y\trans A\trans \Theta_0\inv y \\[0.2cm]
&=& 2 y\trans \Theta_0\inv D \Theta_0\inv y
\end{IEEEeqnarray*}
Using this and  recalling that $b = A\bar{x}_0$ we may write
\begin{IEEEeqnarray*}{rCl}
(A\bar{x} - b)\trans \Theta_0\inv (\bar{x}-\bar{x}_0) &=& (\bar{x} - \bar{x}_0)\trans A\trans \Theta_0\inv (\bar{x}-\bar{x}_0)\\[0.2cm]
&=& (\bar{x} - \bar{x}_0)\trans \Theta_0\inv D \Theta_0\inv (\bar{x}-\bar{x}_0)
\end{IEEEeqnarray*}
Using this result in Eq.~(\ref{eq:gamma1_lin}) and substituting in Eq.~(\ref{eq:PI1_tmp}) we finally obtain Eq.~(\ref{eq:PI1_lin}).

\subsection{Equation for $\Pi_3(t)$}

Let us next compute $\Pi_3$ in Eq.~(\ref{eq:PI3}). It can be written as 
\begin{equation}\label{eq:PI3_tmp}
\Pi_3(t) = \Gamma_2(t) - \Gamma_1(t)
\end{equation}
where $\Gamma_1(t)$ is given by Eq.~(\ref{eq:gamma1}) or (\ref{eq:gamma1_lin}) and
\begin{IEEEeqnarray}{rCl}\label{eq:gamma2}
\Gamma_2 (t) &=& \int \frac{\partial \PP(x,t)}{\partial t} \log \PP_0 (Ex) \ud x \nonumber \\[0.2cm]
                         &=& \int g(x,t)\trans \frac{\partial \PP_0(Ex)}{\partial x} \frac{\ud x}{\PP_0(Ex)}
\end{IEEEeqnarray}
We now have 
\[
\frac{\partial \PP_0(Ex)}{\partial x}  =  - E \Theta_0\inv (Ex - \bar{x}_0)\PP_0 (Ex)
\]
Using again the definitions $F = D\Theta\inv - A$ and $u = b - D\Theta\inv \bar{x}$ we find 
\[
\Gamma_2 = - \int (F x + u)\trans E \Theta_0\inv (Ex - \bar{x}_0) P(x,t) \ud x
\]
Using the method of Eq.~(\ref{eq:ave_method}) we obtain
\[
\Gamma_2 = - \tr(F\trans E \Theta_0 E \theta) - (F\bar{x} + u)\trans E \Theta_0 (E \bar{x} - \bar{x}_0)
\]
The first term reads
\[
-\tr(F\trans E \Theta_0 E \theta)  = \tr (A\trans E\Theta_0\inv E \Theta - \Theta\inv D E \Theta_0\inv E \Theta)
\]
For the last term we may use the cyclic property of the trace to write it as $\tr (EDE\Theta_0\inv)$. However, since $D$ is diagonal, we have that $EDE = D$. Whence, the last term is simply $\tr(D\Theta_0\inv) = \tr(A)$, according to Eq.~(\ref{eq:ent_equi}). 

Collecting our results we finally conclude that 
\begin{IEEEeqnarray}{rCl}\label{eq:gamma2_lin}
\Gamma_2 (t) &=& \tr(A\trans E \Theta_0\inv E \Theta - A)  \nonumber \\[0.2cm]
&+& (A\bar{x}-b)\trans E\Theta_0\inv (E\bar{x}-\bar{x}_0)
\end{IEEEeqnarray}
Using this and Eq.~(\ref{eq:gamma1_lin}) in  Eq.~(\ref{eq:PI3_tmp}) we finally obtain Eq.~(\ref{eq:PI3_lin}).

\subsection{Equation for $\Pi_2(t)$}

Finally, we turn to $\Pi_2$ in Eq.~(\ref{eq:PI2}). We begin by writing 
\begin{IEEEeqnarray}{rCl}\label{eq:gori_lin}
g_0\ir (Ex) &=& \Big[(D\Theta_0\inv E - A\ir E) x + b\ir - D\Theta_0\inv \bar{x}_0\Big] \PP_0(Ex)\nonumber \\[0.2cm]
                  &=&  (Fx + u) \PP_0 (Ex) 
\end{IEEEeqnarray}
We then have 
\begin{IEEEeqnarray}{rCl}\label{eq:PI2_tmp}
\Pi_2 &=& \int (Fx+u)\trans D\inv (Fx+u) \PP(x,t) \ud x \nonumber \\[0.2cm]
	 &=& \tr(F\trans D\inv F \Theta) + (F\bar{x}+u)\trans D\inv (F \bar{x}+u)
\end{IEEEeqnarray}

Let us first simplify the trace term:
\begin{IEEEeqnarray}{rCl}\label{eq:trace_tmp}
\tr (F\trans D\inv F \Theta) &=& \tr\Big[E\Theta_0\inv (D\Theta_0\inv - 2 A\ir) E \Theta\Big]  \nonumber \\[0.2cm]
    					  &+& \tr(E{A\ir}\trans D\inv A\ir E \Theta)
\end{IEEEeqnarray}
The first term can be simplified as follows. First, we note the following relation stemming from Eqs.~(\ref{eq:A_blocks})-(\ref{eq:Air}):
\begin{equation}\label{eq:A_relation}
A\ir = \frac{1}{2} (A + E A E)
\end{equation}
Using this in the first term of Eq.~(\ref{eq:trace_tmp}) we find 
\begin{IEEEeqnarray*}{rCl}
\tr\Big[E\Theta_0\inv (D\Theta_0\inv - 2 A\ir) E \Theta\Big] &=& \tr\Big[E \Theta_0\inv (D\Theta_0\inv - A )E \Theta\Big] \\[0.2cm]
&-& \tr(E\Theta_0\inv E A \Theta)
\end{IEEEeqnarray*}
Similarly to the first term in Eq.~(\ref{eq:gamma2_lin}), the last term in this equation can be written as $\tr(A\trans E \Theta_0\inv E \Theta)$. Moreover, the first term is zero, which may be shown by substituting 
\[
D\Theta_0\inv = \frac{1}{2} (A + \Theta_0 A\trans \Theta_0\inv)
\]
Finally, the last term in Eq.~(\ref{eq:trace_tmp}) can be written as $\tr({A\ir}\trans D\inv A\ir  \Theta)$. This is a direct consequence of the structure of $A\ir$ in Eq.~(\ref{eq:Air}) and the fact that we are assuming $D$ to be diagonal. In fact, for a non-diagonal $D$ this simplification would not occur and this term would not correctly match it's counterpart appearing  in Eq.~(\ref{eq:PI_lin}). 
Thus, we conclude that 
\begin{equation}\label{eq:trace_tmp2}
\tr (F\trans D\inv F \Theta) = \tr({A\ir}\trans D\inv A\ir  \Theta) - \tr(A\trans E \Theta_0\inv E \Theta)
\end{equation}

For the second term in Eq.~(\ref{eq:PI2_tmp}) we now have
\[
F\bar{x} + u = D\Theta_0\inv (E\bar{x} - \bar{x}_0) + b\ir - A\ir E \bar{x}
\]
Expanding the quadratic form, using Eq.~(\ref{eq:A_relation}) and simplifying finally yields the required corresponding term in  Eq.~(\ref{eq:PI2_lin}).

\bibliography{/Users/gtlandi/Documents/library}

\end{document}